\documentclass[journal=ancac3]{achemso}
\setkeys{acs}{keywords = true}

\title{Band Gap Opening by Two-Dimensional Manifestation of Peierls Instability in Graphene}

\author{Sung-Hoon~Lee} \email{sung-hoon.lee@samsung.com}
\author{Hyun-Jong~Chung} \author{Jinseong~Heo} \author{Heejun~Yang} 
\author{Jaikwang~Shin} \author{U-In~Chung} 
\author{Sunae~Seo}
\affiliation{Samsung Advanced Institute of Technology, Yongin 446-712, Korea}
\date{\today}


\begin{document}

\begin{abstract} 

Using first-principles calculations of graphene having high-symmetry distortion or defects, we investigate band gap opening by chiral symmetry breaking, or intervalley mixing, in graphene and show an intuitive picture of understanding the gap opening in terms of local bonding and antibonding hybridizations.
We identify that the gap opening by chiral symmetry breaking in honeycomb lattices is an ideal two-dimensional (2D) extension of the Peierls metal-insulator transition in 1D linear lattices.
We show that the spontaneous Kekule distortion, a 2D version of the Peierls distortion, takes place in biaxially strained graphene, leading to structural failure.
We also show that the gap opening in graphene antidots and armchair nanoribbons, which has been attributed usually to quantum confinement effects, can be understood with the chiral symmetry breaking.

\end{abstract}

\newpage

High carrier mobility makes graphene promising as a channel material for field effect transistors \cite{Novoselov04, Zhang05}. The absence of a band gap in graphene, however, gives rise to poor on-off ratios in the transistor performance \cite{Schwierz10}.  Several approaches to open a band gap have been suggested, such as interactions with substrates \cite{Zhou07}, patterning into nanoribbons \cite{Ezawa06,Brey06,Son06b,Barone06}, quantum dots \cite{Ponomarenko08} or into periodic structures containing carbon vacancies, called antidots \cite{Pedersen08,Kim10,Bai10}, periodic modulation with hydrogen adsorption \cite{Balog10}, or imposing strain\cite{Pereira09a,Choi10,Pereira09b,Guinea09}.  The origin of gap opening in those structures has been usually attributed to sublattice symmetry breaking\cite{Semenoff84}, magnetic effects\cite{Son06b}, or quantum confinement effects\cite{Son06b,Ponomarenko08,Pedersen08,Kim10,Bai10,Geim07}. However, it is unclear how quantum confinement effects open a band gap in graphene. In fact, there are many counterexamples.  Under the time reversal symmetry, \emph{i.e.}, without including magnetic effects, graphene nanoribbons with zigzag edges do not have a band gap despite 1D confinement\cite{Ezawa06,Brey06,Son06b,Barone06}.  Even with armchair edges graphene nanoribbons of certain widths have a zero band gap when lattice distortions at edges are disregarded \cite{Ezawa06,Brey06,Son06b}.  Further, studies on graphene under periodic potentials found that confinement itself cannot open a band gap \cite{Barbier08,Park08}. A clear understanding of the gap opening is necessary to accelerate achieving practically viable gapped graphene.

Graphene has a zero band gap with four half-filled degenerate states at the intrinsic Fermi level \cite{CastroNeto09}. The fourfold degeneracy, consisting of two degenerate states at two nonequivalent Dirac points ($K$ and $K'$), comes from the crystal symmetry of graphene's honeycomb lattice.  Band gap opening in graphene thus implies breaking of the symmetry.  Analytical studies based on effective Hamiltonians \cite{Ajiki99,Chamon00,Hou07,Manes07,Ryu09} have proposed various symmetry-breaking mechanisms of gap opening, including sublattice symmetry breaking, chiral symmetry breaking, spin-orbit coupling, \emph{etc.}  Among those, chiral symmetry breaking, or intervalley mixing, that couples Bloch states at the two Dirac valleys with each other has fundamental importance.  Since the electronic states at the $K$ and $K'$ valleys in graphene represent the massless Dirac fermion spectrum of different chiralities \cite{CastroNeto09} (with the spin being the pseudospin defined in the sublattice space), the gap opening by chiral symmetry breaking corresponds to the mass gap generation of the massless Dirac fermions in quantum electrodynamics (QED) \cite{Nambu61a}.

In this paper, we study the gap opening by chiral symmetry breaking in graphene using first-principles calculations. We examine in detail the electronic wavefunctions of gapped graphene having high-symmetry distortion or defects, and show that gap opening by chiral symmetry breaking can be understood easily in terms of local bonding and antibonding hybridizations. 
Especially we identify that the chiral symmetry breaking in honeycomb lattices \emph{via} electron-lattice coupling is an ideal 2D manifestation of the 1D Peierls metal-insulator transition and show that spontaneous 2D lattice distortion occurs in graphene when biaxial strain is applied, initiating structural failure. Finally we show that the gap opening in graphene antidots and armchair nanoribbons can be understood due to chiral symmetry breaking.

\begin{figure}[] \includegraphics[scale=1.00]{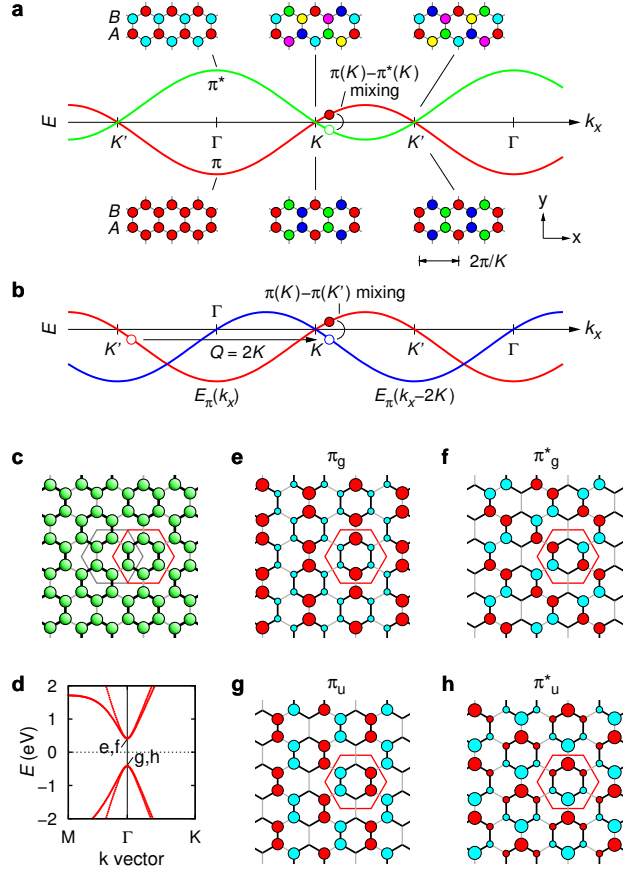} \caption{\label{fig:chiral} 
{Band gap opening by chiral symmetry breaking in graphene's honeycomb lattice.} (a) Band structure of pristine graphene along the line connecting the Brillouin zone center ($\Gamma$) and the $K$ point.  The $\pi(K)$--$\pi^*(K)$ mixing can occur by sublattice-symmetry breaking potentials.  The upper and lower panels show wavefunctions at $\Gamma$, $K$ and $K'$.  The wavefunction coefficients of C $2p_z$ orbitals are reflected in the radius and color of a circle at each atomic site, with the phase angle of 0, $2\pi/3$, and $4\pi/3$ being represented by red, green, and blue, and the intermediate angles by their interpolation.  (b) $\pi(K)$--$\pi(K')$ mixing or chiral symmetry breaking.  A level crossing occurs by the momentum transfer of $\vec Q = \pm 2K \hat x$ by perturbing potentials.  (c) Atomic structure of graphene with the Kekule distortion.  The thick bonds are shorter than the thin bonds.  Two equivalent Wigner-Seitz cells are shown.  (d) Calculated band structure of a Kekule-distorted structure with $\sim 7$ \% bond length asymmetry. (e,f) Wavefunctions of the lowest empty states.  (g,h) Wavefunctions of the highest occupied states.  
} \end{figure}

\subsection{Results and Discussion}

We begin by discussing how gap opening occurs in graphene within time reversal symmetry. In a nearest-neighbor tight-binding model \cite{CastroNeto09}, the electronic structure of pristine graphene is described by two bands, each from the bonding ($\pi$) and antibonding ($\pi^*$) hybridization of two sublattice atoms' $p_z$ orbitals (\ref{fig:chiral}a).  The two bands become degenerate at $K$ and $K'$.  The four degenerate states, $\pi(K)$, $\pi^*(K)$, $\pi(K')$, and $\pi^*(K')$, can undergo energy splitting in two ways \cite{Manes07}, the $\pi$--$\pi^*$ and $K$--$K'$ mixing. The $\pi$--$\pi^*$ mixing or sublattice symmetry breaking occurs by asymmetric on-site energies for the two sublattices, producing dehybridization into sublattice Bloch states, \emph{e.g.}, at the $K$ point, $\psi_{A,B}(\vec K+\vec q) = \psi_\pi(\vec K+\vec q) \pm \psi_{\pi^*}(\vec K+\vec q)$.  The $K$--$K'$ mixing or chiral symmetry breaking occurs by periodic potentials that provide a momentum transfer of $\vec Q = \pm 2\vec K$ (equivalently $\vec Q = \mp \vec K$). Among all four $K$--$K'$ mixing possibilities of $\pi(K)$--$\pi(K')$, $\pi^*(K)$--$\pi^*(K')$, $\pi(K)$--$\pi^*(K')$ and $\pi^*(K)$--$\pi(K')$, the first two play a role in gap opening, because only those mixings introduce level crossing at the intrinsic Fermi level, as depicted in \ref{fig:chiral}b for the $\pi(K)$--$\pi(K')$ mixing.  In an effective $4\times 4$ Hamiltonian acting on a four-component spinor consisting of the four degenerate states, those gap-opening mixing terms correspond to the mass terms in the Dirac equation \cite{Nambu61a,Ajiki99,Chamon00}. They become finite in graphene's honeycomb lattice upon the Kekule distortion \cite{Ajiki99,Chamon00}. The Kekule-distorted structure (\ref{fig:chiral}c) has a Wigner-Seitz cell containing a single benzene unit (the gray hexagon in \ref{fig:chiral}c), where the C--C bonds consist of alternating short and long bonds. It has been studied earlier as a possible low-energy structure for aromatic hydrocarbons \cite{Pauling} and carbon nanotubes \cite{Mintmire92,Okahara94,Ajiki99,Chamon00}, involving the Peierls instability \cite{Peierls} as in 1D atomic chain, but such an instability was found not to occur in 2D graphene at and far below room temperature \cite{Mintmire92,Chamon00}.
(We note that the geometry depicted in \ref{fig:chiral}c is actually the inverse distortion of the typical Kekule distortion in earlier studies\cite{Pauling,Mintmire92,Okahara94,Ajiki99,Chamon00}, for which we will discuss shortly.)

We performed first-principles calculations for a forced Kekule-distorted structure to scrutinize chiral symmetry breaking in graphene.  The calculated band structure (\ref{fig:chiral}d) shows that the distortion breaks the fourfold degeneracy of pristine graphene into two twofold degeneracies, opening a band gap. The wavefunctions of the highest occupied and lowest empty states (\ref{fig:chiral}e-h) reflect the $K$--$K'$ mixing: They are the mixed states of the four degenerate states of pristine graphene, $\psi_{\pi_{g,u}}(\vec q) \equiv \psi_{\pi}(\vec K+\vec q) \pm \psi_{\pi}(\vec K' + \vec q)$ by the $\pi(K)$--$\pi(K')$ mixing and $\psi_{\pi^*_{u,g}}(\vec q) \equiv \psi_{\pi^*}(\vec K+\vec q) \pm \psi_{\pi^*}(\vec K' + \vec q)$ by the $\pi^*(K)$--$\pi^*(K')$ mixing. The wavefunctions of the chirality-mixed states are the standing waves made of planewaves moving along $\vec K$ and $\vec K'$ ($=-\vec K)$, and show a characteristic feature that every C-C pair has a phase difference of 0 and $\pi$, exhibiting apparent local bonding or antibonding hybridization.

The calculated electronic wavefunctions show two interesting features of chiral symmetry breaking in honeycomb lattices. One is that the Kekule distortion in honeycomb lattices is an ideal 2D extension of the 1D Peierls distortion. When we take an alternative Wigner-Seitz cell (the red hexagon in \ref{fig:chiral}c), the distortion shortens all the intracell bonds, while it lengthens all the intercell bonds, as the Peierls distortion does in a two-atom unit cell atomic chain \cite{Hoffmann87}. Further, the gap opening occurs in the same manner in terms of the modification of local bonding and antibonding hybridizations.  Among the four chirality-mixed states $\{\pi_{g,u}, \pi^*_{u,g}\}$ that are degenerate in pristine honeycomb lattices, the $\pi_u$ and $\pi^*_u$ states get a lower energy by the distortion because it enhances bonding hybridizations in the intracell bonds while it lessens antibonding ones in the intercell bonds (\ref{fig:chiral}g,h).  For the $\pi_g$ and $\pi^*_g$ states, on the other hand, the opposite occurs (\ref{fig:chiral}e,f), raising their energy.  We found that there is more correspondence between the 2D honeycomb and 1D linear lattices: Both lattices in nearest-neighbor tight-binding models show massless Dirac fermion spectra with fourfold and twofold degeneracy, respectively, at the intrinsic Fermi level and both undergo two common gap opening mechanisms originating from asymmetry in on-site energies and that in hopping integrals, respectively, each corresponding to sublattice and chiral symmetry breaking (Supporting Information). This indicates that, regarding the electronic structures within tight-binding descriptions, the honeycomb lattice, not a square or rectangular lattice, is the 2D extension of a linear lattice, and to the same extent the Kekule distortion is the 2D extension of the Peierls distortion.

\begin{figure}[t] \includegraphics[scale=1.00]{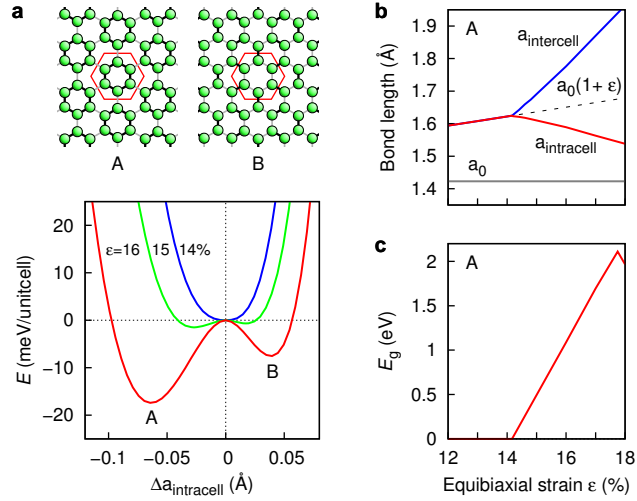} \caption{\label{fig:strain} 
{Spontaneous chiral symmetry breaking in graphene under biaxial strain.}
(a) The energy profile calculated as a function of the relative intracell atomic distances with the equibiaxial strain $\varepsilon = \varepsilon_x = \varepsilon_y$ of 14, 15, and 16\%.  The upper pannels show the schematic atomic geometries of two local minima, $A$ and $B$, corresponding to the inward and outward Kelule distortions.  (b) The intracell and intercell bond lengths for the inward distortion ($A$), as a function of $\varepsilon$.  They bifurcate at $\varepsilon = 14.2 \%$.  $a_0$ is the equilibrium bond length at $\varepsilon = 0$.  (c) The resulting band gap for the inward distortion ($A$).
} \end{figure}

Another interesting feature is the parity symmetry revealed in the chirality-mixed states. According to QED, massive Dirac fermions have an intrinsic parity, pertaining to internal structures of point particles, similar to the intrinsic spin, and their antifermions have an opposite parity \cite{Perkins}. The wavefunctions in \ref{fig:chiral}e-h have parity symmetry and show opposite parities for the occupied ($\pi_u, \pi^*_u$) and empty ($\pi_g, \pi^*_g$) states. The parity symmetry is defined within the Wigner-Seitz cell having six atoms inside, and each of the Wigner-Seitz cells that constitute their own hexagonal lattice has a constant wavefunction amplitude throughout the 2D space. The wavefunctions therefore visualize the product of the parity wavefunction and the spatial wavefunction of massive Dirac fermions at rest in a hexagonal lattice. The two degenerate states for the empty (occupied) states correspond to different intrinsic spin states of the massive Dirac fermions (antifermions); The spin states are seen as the relative phases between sublattice atoms in the wavefunctions (\emph{i.e.}, the pseudospin for graphene). Thus the present results show that the internal structure of 2D point particles in a continuum description is visualized in a lattice description as a six-site internal structure of a hexagonal lattice. 

\begin{figure}[t] \includegraphics[scale=1.00]{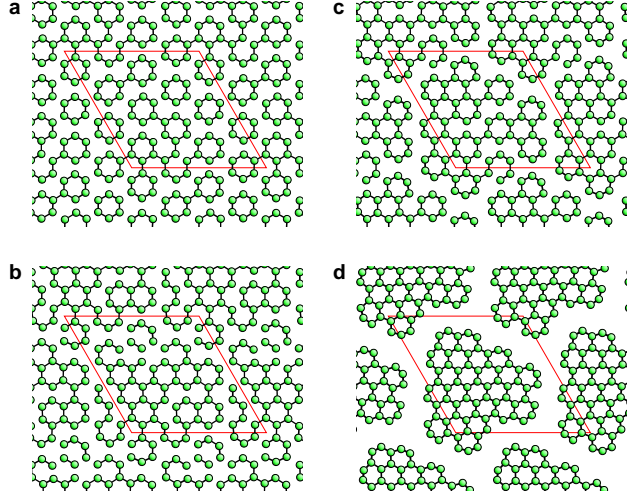} \caption{ \label{fig:MD} 
{Tearing of a graphene sheet \emph{via} spontaneous Kekule distortion under biaxial strain.}
First-principles molecular-dynamics calculations are performed for graphene under an equibiaxial strain $\varepsilon = 15.5\%$ at $T = 300$ K. A $(6\times6)$ supercell is used (red rhombus in the figures).
(a) $t = 20$ fs. With the applied strain, graphene undergoes a prompt Kekule distortion.
(b) $t = 50$ fs. By thermal fluctuation, some longer bonds appear and trigger weakening of neighboring long bonds.
(c) $t = 100$ fs. Boundaries made of longer bonds become prominent.
(d) $t = 300$ fs. Each domain of graphene contracts to relieve the tensile stress and recover equilibrium atomic distances.
} \end{figure}

The band gap opening by the symmetry-breaking distortion that normally does not occur in 2D graphene takes place spontaneously in biaxially strained graphene.
Our first-principles calculations show that a spontaneous Kekule distortion occurs at equibiaxial strain of $\varepsilon = \varepsilon_x= \varepsilon_y > 14\%$ (\ref{fig:strain}a,b). It shows asymmetric two local minima, the inward ($A$) and outward ($B$) distortions. The outward Kekule distortion leading to dimerization of carbon atoms, which has been discussed previously \cite{Pauling,Mintmire92,Okahara94,Ajiki99,Chamon00} as a 2D extension of the Peierls distortion, is less stable than the inward distortion. 
The preference to the inward distortion of graphene comes from the $\sigma$ bonds in graphene: We performed calculations of a comparative honeycomb lattice made of hydrogen atoms and found that it favors the outward distortion (Supporting Information).
The calculations also indicate that the strong $\sigma$ bonds are responsible for the lack of the Kekule distortion in graphene at zero strain.
The distortion under biaxial strain produces a finite band gap (\ref{fig:strain}c) that grows rapidly with increasing strain after the onset at $\varepsilon = 14.2\%$ up to $\varepsilon = 17.8\%$ where the $\sigma^*$ band starts to descend below the empty $\pi$ band. This chiral-symmetry-broken state of graphene, however, is subject to a structural failure triggered off at fluctuating long bonds: Our molecular dynamics simulations of graphene under a biaxial strain of 15.5\% result in tearing of the graphene sheet \emph{via} the Kekule distortion (\ref{fig:MD}). We note that the symmetry breaking and gap opening under biaxial strain is distinct from that by uniaxial strain \cite{Pereira09a,Choi10}, where the gap opening occurs by the gradual merging of the two Dirac points caused by the uniaxial distortion of the honeycomb lattice \cite{Pereira09a}.

\begin{figure}[t] \includegraphics[scale=1.00]{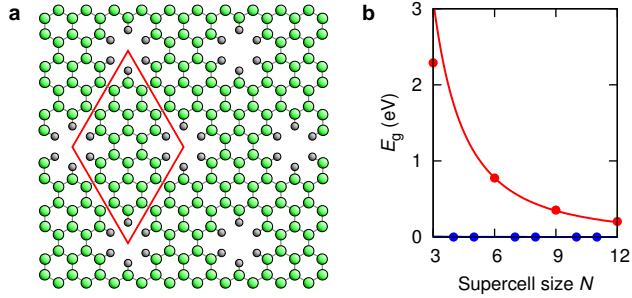} \caption{\label{fig:Eg} {Band gap opening in graphene by hexagonal antidot superlattices.} (a) Atomic structure for the $(4\times 4)$ supercell ($N = 4$). The antidot consists of six carbon vacancies in the shape of benzene ring. Dangling bonds are passivated by hydrogen atoms so that the $sp^2$ bond network is intact. (b) Calculated band gaps as a function of the antidot supercell size ($N$). Red and blue dots represent different characteristic of the data. Solid lines are fitted to $N^{-2}$, proportional to the density of defects. } \end{figure}

\begin{figure}[t] \includegraphics[scale=1.00]{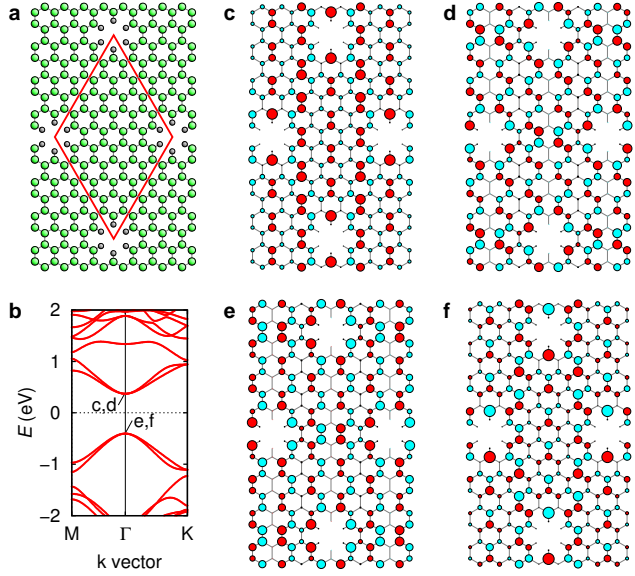} \caption{\label{fig:wf} {Explicit chiral symmetry breaking in graphene by hexagonal antidot superlattices.} (a) Atomic structure for the $(6\times6)$ supercell.  (b) Calculated band structure.  For the $(6\times 6)$ supercell, the two Dirac points fold into the $\Gamma$ point in the reduced Brillouin zone.  The fourfold degeneracy is broken into two twofold degeneracies.  (c,d) Wavefunctions of the lowest empty states, each corresponding to $\pi_g$ and $\pi_g^*$.  (e,f) Wavefunctions of the highest occupied states, each corresponding to $\pi_u$ and $\pi_u^*$.
} \end{figure}

The chiral symmetry breaking in graphene can occur not only by spontaneous lattice distortions but also by explicit lattice defects. We now show the manifestation of chiral symmetry breaking in previously reported structures of gapped graphene, for which gap opening was attributed to quantum confinement effects or others.  We consider the periodic antidot defects first\cite{Pedersen08,Kim10,Bai10}. Hexagonal antidot superlattices with antidots composed of six carbon vacancies (\ref{fig:Eg}a) reduce the translational symmetry, yet preserving graphene's $C_{6v}$ point group symmetry.  The calculated band gaps for the $(N\times N)$ supercell (\ref{fig:Eg}b) show that they are nonzero and proportional to the density of defects only when the size of the antidot superlattice, $N$, is a multiple of 3.  This supercell-size dependence invalidates quantum confinement effects as the origin of gap opening.  The calculated wavefunctions (\ref{fig:wf}) show that chiral symmetry breaking instead is the origin: The lowest empty and highest occupied states represent nothing but the chirality-mixed states, $\{\pi_g, \pi^*_g\}$ and $\{\pi_u, \pi^*_u\}$, respectively.  The gap opening, or the degeneracy lifting of the four chirality-mixed states, can be understood as the result of a systematic removal of bonding neighbors for $\{\pi_g, \pi^*_g\}$ and of antibonding neighbors for $\{\pi_u, \pi^*_u\}$ by the antidot formation. This gap opening depends on the commensurability of the chirality-mixed states with the defect lattice: When they are incommensurate, \emph{i.e.}, when $N$ is not a multiple of 3, the antidot formation removes both bonding and antibonding neighbors, resulting in a net energy change of zero, for all the chirality-mixed states.

\begin{figure}[t] \includegraphics[scale=1.00]{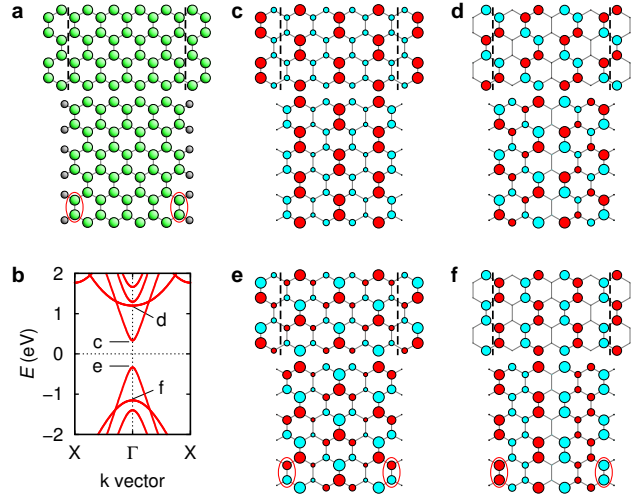} \caption{ \label{fig:NR9}
{Band gap opening by chiral symmetry breaking in armchair graphene nanoribbons.}
(a) Atomic structure of an armchair nanoribbon with nine carbon dimers along the width.
(b) Calculated band structure. The fourfold degeneracy is fully lifted.
(c,d) Wavefunctions of the lowest and second lowest empty states at $\Gamma$, corresponding to $\pi_g$ and $\pi^*_g$.
(e,f) Wavefunctions of the highest and second highest occupied states at $\Gamma$, corresponding to $\pi^*_u$ and $\pi_u$.
The edge truncation, depicted in the upper panel, makes the $\pi^*_g$ and $\pi_u$ states deviate significantly from zero energy and thus mix strongly with high-energy states.
The edge dimer length is shortened spontaneously by $3.5$ \% compared to that in bulk \cite{Son06b}.  It is the consequence of the stronger bonding hybridization between the edge dimer atoms of the second highest occupied state, compared to the antibonding hybridization of the highest occupied state.  The edge dimer distortion thus reduces the band gap slightly.  
} \end{figure}

Similarly one can understand the gap opening in quasi-1D graphene nanoribbons with armchair edges \cite{Ezawa06, Brey06, Son06b, Barone06}, where the armchair edges impose the chiral symmetry breaking. As shown in \ref{fig:NR9}, the edge truncation removes bonding neighbors for $\pi_g$ and $\pi^*_g$, while it removes antibonding ones for $\pi^*_u$ and $\pi_u$, and thereby lifts the degeneracy among them. For the $\pi_g$ and $\pi^*_u$ states (\ref{fig:NR9}c,e) that have weak hybridizations at the edges, the energy change is small, but for the $\pi^*_g$ and $\pi_u$ states (\ref{fig:NR9}d,f) that have strong hybridizations at the edges, the energy change is large, leading to strong mixing with high-energy states.
Similar electronic structures can be produced in 2D graphene with periodic line patterns of hydrogen adsorption along the armchair direction (Supporting Information), where the passivation of $p_z$ orbitals by hydrogen adsorption plays the role of edges.
The inverse proportionality of the band gap to the ribbon width, which has been regarded as an evidence for the manifestation of quantum confinement effects for armchair nanoribbons\cite{Son06b, Geim07}, can thus be understood as just the result of the decreasing density of chiral symmetry breaking defects as the width increases.

\begin{figure}[t] \includegraphics[scale=1.00]{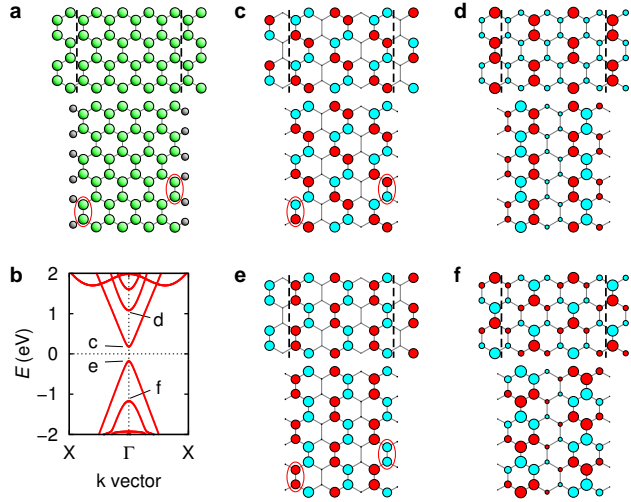} \caption{\label{fig:NR8} 
{Band gap opening by chiral symmetry breaking in armchair graphene nanoribbons.}
(a) Atomic structure of an armchair nanoribbon with eight carbon dimers along the width.
(b) Calculated band structure.  The fourfold degeneracy is fully lifted.
(c,d) Wavefunctions of the lowest and second lowest empty states at $\Gamma$, corresponding to $\pi^*_g$ and $\pi_g$.
(e,f) Wavefunctions of the highest and second highest occupied states at $\Gamma$, corresponding to $\pi_u$ and $\pi^*_u$.
The edge truncation, depicted in the upper panel, makes the $\pi_g$ and $\pi^*_u$ states deviate significantly from zero energy and thus mix strongly with high-energy states. 
Here the $\pi_g$ state gets a higher energy despite the removal of antibonding neighbors for the edge atoms. It is because that the edge truncation also removes strong bonding pairs that make a higher contribution to the energy. Similarly the $\pi^*_u$ state gets a lower energy.
For the $\pi^*_g$ and $\pi_u$ states the edge truncation removes just nonbonding neighbors and their degeneracy is lifted by spontaneous distortion of edge dimers.
} \end{figure}

An interesting case arises when the armchair nanoribbons have a certain width of $(3n-1)$ carbon dimers; the $n=3$ case is shown in \ref{fig:NR8}. With these widths, the $\pi_g$ and $\pi^*_u$ states (\ref{fig:NR8}d,f) undergo large energy splitting by the edge truncation, but the $\pi^*_g$ and $\pi_u$ states (\ref{fig:NR8}c,e) are yet degenerate at zero energy \cite{Ezawa06,Brey06}, because they have nodal lines along the armchair edges and the edge truncation removes just nonbonding neighbors at both edges. The degeneracy is lifted actually by spontaneous dimer distortion at edges leading to 3.5 \% reduction in the bond length \cite{Son06b}. This so-called edge effect is again the manifestation of chiral symmetry breaking: The edge distortion enhances bonding hybridizations for $\pi_u$, whereas it enhances antibonding hybridizations for $\pi^*_g$, leading to the energy splitting between them. This spontaneous gap opening, allowed in graphene for the reduced elastic energy cost from the low symmetry at edges, is identical to the 1D Peierls transition.  It is especially true for the $n=1$ case which is just the 1D atomic chain system of \emph{cis}-polyacetylene \cite{Su80}, with only the $\pi^*_g$ and $\pi_u$ states being involved for the gap opening. This demonstrates that the gap opening by chiral symmetry breaking in graphene is indeed a 2D superset of the 1D Peierls transition.

\subsection{Conclusion}

We have shown using detailed analysis of the electronic wavefunctions of gapped graphene that the gap opening by chiral symmetry breaking in graphene can be understood as an ideal 2D superset of the 1D Peierls transition and also as the degeneracy lifting of four-chirality-mixed states, which are represented as the network of local bonding and antibonding hybridizations. 
Our study have shown that this understanding is useful, providing a simple, unified description of the gap opening in 2D graphene antidots and quasi-1D armchair nanoribbons and predicting structural failure of biaxially strained graphene \emph{via} spontaneous 2D lattice distortion.

\subsection{Methods}

Our first-principles calculations are based on the density functional theory (DFT) employing the generalized gradient approximation and the projector-augmented-wave method as implemented in VASP\cite{Kresse96,Kresse99}.  Valence electronic wavefunctions are expanded in a planewave basis set with a cutoff energy of 280 eV. In our supercell calculations, graphene layers are separated from each other by 8 \AA. The $k$-point integration was made at a uniform $k$-point mesh of ($30 \times 30$) per unit cell. The atomic positions are relaxed until residual forces are less than 0.02 eV/\AA.

\begin{acknowledgement}

The authors thank Hyoung Joon Choi and A.~H.~Castro Neto for discussions and comments.

\end{acknowledgement}

\begin{suppinfo}

Additional figures regarding the gap opening in 1D atomic chains, the 2D Kekule distortion in hydrogenic honeycomb lattices, and the electronic structure of graphene with periodic line patterns of hydrogen adsorption.

\end{suppinfo}

\newpage
\providecommand*{\mcitethebibliography}{\thebibliography}
\csname @ifundefined\endcsname{endmcitethebibliography}
{\let\endmcitethebibliography\endthebibliography}{}

\newpage

\end{document}